# Boron adatom adsorption on graphene: A case study in computational chemistry methods for surface interactions

**Sierra Jubin[1,2*], Aaditya Rau[3], Yuri Barsukov[1], Stephane Ethier[1], Igor Kaganovich[1]**

[1]Princeton Plasma Physics Laboratory (PPPL), Princeton, NJ, USA
[2]Princeton University, Princeton, NJ, USA
[3]Johns Hopkins University, Baltimore, MD, USA

*** Correspondence:**
Sierra Jubin
sjubin@princeton.edu

**Abstract**

Though weak surface interactions and adsorption can play an important role in plasma processing and materials science, they are not necessarily simple to model. A boron adatom adsorbed on a graphene sheet serves as a case study for how carefully one must select the correct technique from a toolbox of computational chemistry methods. Using a variety of molecular dynamics potentials and density functional theory functionals, we evaluate the adsorption energy, investigate barriers to adsorption and migration, calculate corresponding reaction rates, and show that a surprisingly high level of theory may be necessary to verify that the system is described correctly.

## 1 Introduction

Examining the adsorption of atomic boron on a graphene sheet offers insight into the difficulties of modelling surface interactions, which can play an important role in materials processing applications. Our interest in using classical molecular dynamics to model the process by which boron is deposited on graphite plasma facing components in fusion devices led us to wonder how boron adheres to the surface of a polycrystalline piece of graphite, though admittedly these adsorption processes occur at far lower energies than those relevant for a fusion device. A wide variety of computational approaches are available to investigate adsorption processes, including classical molecular dynamics (MD), density functional theory (DFT) methods, post-Hartree-Fock methods, and multi-configurational methods. It can be difficult to know ahead of time the right balance between computational cost and the level of accuracy required by the problem under study. Thus, understanding the level of theory necessary to correctly identify binding sites and surface chemistry is important, especially for weak interactions – such as the adsorption of atomic boron on a graphene sheet.

It is worth noting that in addition to general interest as an avenue for investigating methods of modelling surface adsorption, the structure of atomic boron adsorbed on graphene has a variety of suggested applications. It has been proposed as an intermediate structure in a mechanism for the barrier-free substitutional doping of graphene with boron atoms, a material which can act as a sensor for carbon monoxide gas, and a structure with a localized magnetic moment. [1-3] This last application was investigated by Li *et al* in a study in which the structure was both modeled using DFT calculations and observed experimentally and probed with x-ray photoelectron spectroscopy. [3]

Other studies of atomic boron adsorbed onto graphene have used only DFT methods to identify the adsorption site and energy. [1-2,4-7]
To illustrate the difficulty of achieving some reasonable approximation of chemical accuracy when investigating weak surface interactions such as the adsorption of boron on graphene surfaces, we performed calculations using molecular dynamics (MD) potentials, DFT methods with a variety functionals, and second order Moller-Plesset (MP2) methods.

## 2 Computational methods

There are three possible binding sites for adatoms on the surface of a graphene sheet: the top site (T-site) directly above a carbon atom, the bridge site (B-site) directly over the bond between two carbon atoms, and the hollow site (H-site) centered over a hexagon. (See **Figure 1**.) Previous DFT studies have found that for a boron adatom, the B-site is the most favorable. [1-7] Calculated adsorption energies and distances between boron and the nearest carbon atom are shown in **Table 1**.
Though all previous studies agree that the B-site is the most energetically favorable, the discrepancy in the calculated binding energy is quite striking. The values given range from 0.24 to 1.8 eV, though the distance between the boron atom and nearest carbon atoms of the stable structure, when given, was calculated to be ~1.8 Å in each of the previous studies. [1-7] Notably, many of these calculations were implemented using the PBE functional of Perdew, Burke and Ernzerhof, with a handful of exceptions. [8] To our knowledge, the only calculation of the energy barrier for the migration of boron adatoms between adjacent binding sites was done by Nakada and Ishii using an LDA functional. They estimated this migration barrier to be 0.12 eV. [5]

### 2.1 Classical molecular dynamics methods

While typically less accurate than quantum chemistry methods, classical molecular dynamics (CMD) allows for the simulation of much larger systems over much larger time scales due to its lower computational cost. Interatomic potentials are used to calculate the forces on particles, and the equations of motion are numerically solved to track the trajectory of the particles over time. For simulations that involve accurately modelling bond breaking and formation, bond order potentials (BOP) enable the description of different bonding states between atoms. These interatomic potentials are produced by parameterizing empirical formulae based on data often produced via quantum chemistry calculations, such as DFT descriptions of bond lengths, bond angles, and the potential energy surface near pertinent molecular geometries. Naively, one might assume that a MD potential optimized on molecules and structures sufficiently similar to the relevant surface interaction would yield a reasonable estimate for the adsorption energy. MD potentials are regularly used to simulate reactions that are not identical to the reactions on which they were fitted. Though this may work well in some situations, for applications involving weak interactions such as the adsorption of boron on graphene, the fidelity of these potentials requires validation.
Our CMD simulations were run using the Large-scale Atomic/Molecular Massively Parallel Simulator (LAMMPS) code. [9] In order to model the adsorption of boron adatoms on graphene, we investigated three different BOPs. These include a Tersoff-type potential and three ReaxFF potentials. [10-13] This particular Tersoff-type potential was developed to investigate thermal conductivity in hexagonal boron nitride and graphene hybrid nanostructures and was parameterized to reproduce DFT energetics obtained using the PBE functional – specifically, the energy as a function of interface distance between hexagonal BN and graphene sheets. [10] The ReaxFF potentials belong to a family of bond order potentials first developed in 2001 and modified to its current form in 2008. [14,15] The first ReaxFF potential (ReaxFF-1) was developed to investigate deuterium uptake in amorphous carbon surfaces containing varying amounts of lithium, boron, and carbon, and it is a modified form of a potential initially developed to model chemistry at the anode-electrolyte interface of lithium-sulfur batteries. [11, 16] The data used for parameterization includes DFT energetics of lithium interacting with a boron substituted cyclohexane ring. [11] The second ReaxFF potential (ReaxFF-2) was developed to model shear deformations of boron carbide, and like the Tersoff-type potential, it was also parameterized to match DFT calculations obtained using the PBE functional. The parameters were fitted so that the potential matched data regarding the interaction of two $B_{10}C_2H_{12}$ icosahedra, the equations of state and heats of formation for various phases of boron and boron carbide, and the shear deformation of boron carbide. [12] The third and final ReaxFF potential (ReaxFF-3) was developed to model liquid CBN (carbon-boron-nitride) hydrogen-storage materials, parameterized by fitting DFT calculations of potential energy surfaces related to varying bond lengths/angles for a variety of small molecules containing carbon, hydrogen, and/or boron. [13] This last ReaxFF potential is in fact part of the 'aqueous branch' of the ReaxFF development tree, whereas the first two are from the 'combustion branch.' [17] In conjunction with the ReaxFF implementation, the charge equilibration (QEq) method was used to compute the charges on each atom at every timestep.[18]

We prepared the initial geometry of the graphene sheet by constructing a bilayer graphite crystal with planar dimensions of approximately 40 Å by 35 Å and periodic boundary conditions. After allowing the system to relax to a minimized energy (including the dimensions of the simulation cell) the top layer of graphene was deleted, leaving a single layer of graphene consisting of 512 atoms. This is the equivalent area of 16 x 16 unit cells of graphene. We used the resulting geometry to investigate graphene-boron surface interactions as the carbon atoms of the graphene sheet were held stationary. For each interatomic potential and each of the three binding sites specified in **Figure 1**, a single boron atom was moved from 4.5 Å above the binding site towards the graphene sheet and the potential energy of the system was recorded along this trajectory. Minima along each one-dimensional potential energy scan were compared to identify the most stable binding site for each MD potential.

## 2.2 Quantum chemistry methods

All quantum chemistry calculations were carried out using cluster models in Gaussian 16 software. [19] Large planar aromatic hydrocarbons – bisanthene ($C_{28}H_{14}$) and coronene ($C_{24}H_{12}$) – were used as a model for graphene. Such polyaromatic hydrocarbons have previously been used as a model for graphene in computational investigations of adsorption on graphene surfaces. [20-21] Though adsorption energies calculated using finite models may differ from adsorption energies calculated for an infinite graphite sheet represented by periodic boundary conditions, we expect qualitative agreement regarding the predicted stable geometries. Binding energies calculated for methane interacting with coronene have previously been shown to be within a few tenths of an eV to the binding energies predicted for methane interacting with much larger polyaromatic hydrocarbons (consisting of hundreds of atoms), indicating that coronene is sufficiently large to qualitatively model adsorption on infinite graphene sheets. [22]

In addition to adsorption on coronene and bisanthene, we explored the interactions of atomic boron with benzene ($C_6H_6$). These calculations were performed mainly for comparison with the calculations previously done by Bettinger and Kaiser using multiconfigurational methods to study the formation of benzoborirene. [23] Though benzene is too small to be a good model for an infinite graphene sheet, its interactions with boron serve as a way to compare the performance of various DFT functionals against more accurate calculations performed at a higher level of theory. We used three different DFT functionals of the generalized gradient approximation (GGA) form, including the PBE functional described earlier in this work, Becke's three-parameter hybrid exchange functional (B3LYP), and a hybrid exchange functional with empirical long-range dispersion (wB97XD). [8, 24-25] We employed these functionals in conjunction with the double zeta correlation-consistent Dunning basis set, cc-pVDZ. [26] The default convergence criteria of the Gaussian 16 code were implemented – namely, energy convergence is reached if the energy difference between two self consistent field (SCF) cycles is less than $10^{-6}$ Hartree. Adsorption energies were calculated as below:

$$\boldsymbol{E_{ads} = E_{C_xH_yB} - E_{C_xH_y} - E_B} \quad (1)$$

Where $E_{C_xH_yB}$ represents the electronic energy of the bound structure, $E_{C_xH_y}$ represents the electronic energy of the hydrocarbon alone, and $E_B$ the electronic energy of a lone boron atom. Energy barriers are calculated similarly, as the difference between the transition state energy and the energy of the initial configuration, whether that be separate products or the adsorbed state. All energies have been corrected by the zero-point vibrational energy.
Transition states corresponding to barriers to adsorption and barriers to migration between adjacent binding sites were investigated as well. These transition states correspond to saddle points in the potential energy landscape, and are identified by the presence of one imaginary frequency in the vibrational spectrum. Reaction rates were calculated from the associated energy barriers as a function of temperature, T:

$$\boldsymbol{k_{ads} = \frac{k_B T}{h} \frac{Z_{vib}^{TS}}{Z_{vib}^{C_xH_y} Z_{tot}^{B}} e^{-E_{bar}/k_B T}} \quad (2)$$

$$\boldsymbol{k_{mig} = \frac{k_B T}{h} \frac{Z_{vib}^{TS}}{Z_{vib}^{C_xH_yB}} e^{-E_{bar}/k_B T}} \quad (3)$$

Here, $\boldsymbol{k_B}$ is the Boltzmann constant and $\boldsymbol{h}$ is Planck's constant. $\boldsymbol{Z_{vib}^{TS}}$, $\boldsymbol{Z_{vib}^{C_xH_y}}$, $\boldsymbol{Z_{tot}^{B}}$, and $\boldsymbol{Z_{vib}^{C_xH_yB}}$ are the partition functions associated with the vibrational energy of the relevant transition state, the vibrational energy of the hydrocarbon alone, the total energy of the boron atom, and the vibrational energy of the bound state. $\mathbf{E_{bar}}$ is the height of the energy barrier for the relevant process.

## 3 Results and Discussion

### 3.1 Classical molecular dynamics results and discussion

The potential energy curves are presented in **Figure 2**, scanning over the distance between the boron atom and the graphene sheet. For each MD potential, the preferred binding site and corresponding binding energy minimum are presented in **Table 2**. The adsorption energies predicted by the ReaxFF potentials are larger than any of those predicted by DFT methods in **Table 1**, and the adsorption energy of the Tersoff-type potential is lower. Examining the potential energy curves for the ReaxFF potentials we find large and unphysical fluctuations in the potential energy as the boron atom moves closer to the graphene sheet. The Tersoff-type potential and two of the ReaxFF potentials predict that the hollow site is the most stable, whereas the other ReaxFF potential predicts that the bridge site is the most energetically favorable position for the boron adatom.
None of these MD potentials were parameterized to reproduce the energetics of a boron adatom adsorbing on a graphene sheet, and their unsuitability for this task is not a mark against them. The systems used to parameterize the potentials often did not remotely resemble graphene, with the exception of the Tersoff-type potential. However, we can see that even the Tersoff-type potential predicted a binding energy that fell outside the broad range of binding energies predicted by previous quantum chemistry calculations. The two potentials that were optimized based on DFT calculations for smaller molecules (ReaxFF-1 and ReaxFF-3) predicted unreasonably large adsorption energies, whereas the potentials that were optimized based on DFT calculations for systems containing slabs of graphene and boron carbide (Tersoff-type and ReaxFF-2, respectively) were smaller and more reasonable. Still, the potential energy scans sufficiently demonstrate that the investigated carbon-boron bond order potentials are unsuitable for reproducing the surface interactions of boron adatoms adsorbing on graphene, and one should not assume that molecular dynamics potentials will be capable of accurately modelling surface interactions if they have not been optimized using DFT calculations specific to those reactions.

## 3.2 Quantum chemistry results and discussion

### 3.2.1 Large planar hydrocarbons interacting with boron

It is immediately apparent from the optimized geometries shown in **Figure 3** that the PBE functional predicts a different binding site than the other two functionals. While B3LYP and wB97XD predict that the boron atom will bind near a top site, canted slightly towards a carbon-carbon bond, the PBE functional predicts that the boron atom will bind at the bridge position, as it has in previous studies. The predicted adsorption energies for boron on bisanthene using the B3LYP, wB97XD, and PBE functionals were -0.72 eV, -1.02 eV, and -1.11 eV, respectively. (See **Table 3**.) These fall within the range of previously predicted adsorption energies of boron on graphene. (See **Table 1**.)

Transition states representing barriers to adsorption were identified in for the B3LYP and wB97XD functionals, in addition to a weakly physisorbed state at boron-carbon distances beyond the transition state. Potential energy scans following the intrinsic reaction coordinate path from the weakly physisorbed state, over the transition state, and to the fully adsorbed state are depicted in **Figure 4**. When zero-point energy (ZPE) corrections are taken into account, the transition state energy predicted with the B3LYP method actually lies below the energy of the separated products. It does not, therefore, pose a barrier to adsorption. In contrast the ZPE-corrected barrier predicted by wB97XD is 0.16 eV. Calculations using the PBE functional located no equivalent transition state, as the potential energy was found to monotonically increase as the boron atom was pulled farther from the surface. Thus, only the wB97XD functional predicts the existence of a barrier to adsorption.

Diffusion along the graphene sheet was also considered. In migrating between adjacent binding sites, the boron adatom experiences a barrier to diffusion. This migration was modeled using bisanthene for the B3LYP and wB97XD functionals and coronene for the PBE functional. The predicted barriers to diffusion are 0.20 eV for wB97XD, 0.11 eV for B3LYP, and 0.09 eV for PBE. The potential energy scans along the intrinsic reaction coordinates for this migration process can be seen in **Figure 5**.

The reaction rates calculated from the variety of energy barriers described previously are plotted as a function of temperature in **Figure 6**. A summary of the various calculated energies (adsorption energies, barriers to adsorption, and barriers to migration) with and without zero-point energy corrections can be found in **Table 3**.

### 3.2.2 Benzene interacting with boron

Given the discrepancies found above, it seems that we must inevitably conclude that either the PBE functional or both the B3LYP and wB97XD functional incorrectly predict the binding site for boron adsorption on graphene. One would tentatively expect the hybrid functionals to be more accurate than the PBE functional in predicting reaction barriers and binding energies, existing on a “higher rung” of the DFT functional “Jacob’s ladder” to heavenly chemical accuracy, the popular analogy first used by Perdew. [27] However, for further assurance we turn to modeling the interaction of atomic boron with a smaller aromatic planar hydrocarbon – benzene. Though the interaction between benzene and atomic boron is a crude model for the interaction between graphene and atomic boron, the six-membered carbon ring of benzene is nevertheless a sufficient proxy for the six-membered ring of graphene. Both benzene and graphene are aromatic compounds with delocalized electrons shared among pi bonds; the main difference is that the conjugated system of a benzene ring is terminated by hydrogen atoms, whereas in graphene the conjugated system of a six-membered ring is part of the graphene sheet. Therefore, we can use this very small system to compare the results of different DFT functionals to more accurate (and more computationally expensive) calculations done at a higher level of theory.

This comparison will enable us to make reasonable predictions about which DFT functionals are producing accurate results for the interactions of large polyaromatic hydrocarbons with atomic boron. We once again see that the PBE functional predicts a stable structure with the boron atom positioned directly over the bond between two carbon atoms, whereas the B3LYP and wB97XD functionals both predict a stable geometry with the boron atom nearer to one of the carbon atoms. (See **Figure 7**.) These geometries can be directly compared to structures described by Bettinger and Kaiser, who used B3LYP and a complete active space (CAS) method to investigate the formation of benzoborirene. [20] Despite using a different basis set, the stable geometry predicted by their B3LYP method agrees with the stable geometry predicted by our B3LYP method. More importantly, the CAS calculations of Bettinger and Kaiser indicate that a stable structure exists when the boron is directly over one of the carbon atoms (analogous to T-site adsorption on graphene), and that the geometry predicted to be stable by the PBE functional (labeled **5** by Bettinger and Kaiser) is actually a transition state.

As the multi-configurational complete active space self-consistent field (CASSCF) method used in this paper is at a much higher level of theory than any of our density functional theory methods, this indicates that the B3LYP and wB97XD hybrid functionals predict the binding geometry with greater accuracy than the PBE functional. Since many studies of adatom adsorption on graphene have made use of the PBE functional, further investigation of these structures using hybrid functionals such as B3LYP and wB97XD may produce better estimates of adsorption energy and possibly predict different binding sites than those predicted in previous literature.

Interestingly, MP2 calculations did not agree with the CASSCF results, as this post-Hartree-Fock method agreed with the PBE functional in its prediction that the structure shown in **Figure 7(B)** was stable. Typically, the MP2 method is used to validate DFT results. [28] However, it has been shown elsewhere that the interaction of a beryllium atom with a benzene ring involves multi-reference electronic states which MP2 fails to account for, and the MP2 method overestimates the adsorption energy of this system. [29] We might reasonably anticipate that the MP2 method also overestimates the attractive interactions between a boron atom and a benzene ring, and we would generally expect MP2 to be less accurate than the more sophisticated CAS methods used by Bettinger and Kaiser. [23]

## 4 Conclusions

We have investigated the adsorption of atomic boron on graphene, as modeled by bond order MD potentials, and as modeled by large planar hydrocarbons using a variety of DFT functionals. The MD potentials considered here are not suitable for modeling the adsorption of a boron adatom on a graphene sheet, as one might expect considering that they were not parameterized with this interaction in mind. However, the very unphysical oscillations in some of the ReaxFF potentials and the magnitude of the binding energies predicted should serve as a reminder that MD potentials used to investigate processes where weak surface interactions play a role ought to be optimized to reproduce the energetics of the relevant reactions. To our knowledge, there is no such potential for modeling boron adatom adsorption on graphene. In the future, MD potentials that harness machine learning via the SNAP and DeePMD formalisms may be worth exploring for modelling such surface interactions. [30,31]
Three different DFT functionals were used to estimate binding energies, barriers to adsorption, and barriers to migration between adjacent binding sites for boron adatoms on graphene. For the predicted energy barriers, corresponding reaction rates were calculated. We suggest that the bridge binding site predicted by previous studies may be incorrect, and that the true binding site may be the top site. If this is the case, then the PBE functional may yield the wrong adsorption site, and this is important to note not only for its own sake but for the future parameterization of MD potentials dealing with adsorption on surfaces, which rely on DFT calculations to provide a sufficiently realistic description of relevant potential energy surfaces. Care must be taken to correctly assess the binding energies and barriers related to weak surface interactions, and a surprisingly high level of theory may be needed to verify the correct binding site.

## 5 Conflict of Interest

*The authors declare that the research was conducted in the absence of any commercial or financial relationships that could be construed as a potential conflict of interest.*

## 6 Author Contributions

All authors contributed to the conception of this study. SJ ran the quantum chemistry calculations and wrote the first full draft of the manuscript. AR ran the classical molecular dynamics calculations and wrote the first draft of the corresponding sections of the manuscript. YB provided insight on quantum chemistry calculations. SE helped with installation of software and compiling the codes. IDK organized the research effort and managed group interactions. All authors contributed to manuscript revision and approved the submitted version.

## 7 Funding

The research described in this paper was conducted under the Laboratory Directed Research and Development (LDRD) Program at Princeton Plasma Physics Laboratory, a national laboratory operated by Princeton University for the U.S. Department of Energy under Prime Contract No. DE-AC02-09CH11466. The United States Government retains a non-exclusive, paid-up, irrevocable, world-wide license to publish or reproduce the published form of this manuscript, or allow others to do so, for United States Government purposes.

## 8 Acknowledgments

This research used resources of the National Energy Research Scientific Computing Center (NERSC), a U.S. Department of Energy Office of Science User Facility located at Lawrence Berkeley National Laboratory, operated under Contract No. DE-AC02-05CH11231.
In addition, this research used computing resources on the Princeton University Adroit Cluster and PPPL cluster.

**Table 1: Adsorption energies ($E_{ads}$) and nearest boron-carbon distances ($d_{B-C}$) for past DFT studies of atomic boron adsorbed on graphene and graphene-like surfaces. Distances marked with an * were inferred assuming a graphene C-C bond distance of 1.42 Å.**

| | Method | $E_{ads}$ (eV) | $d_{B-C}$ (Å) |
|---|---|---|---|
| **[4]** | System: Unspecified number of unit cells of graphene<br>Functional: PW91<br>Basis set: planewave | 1.27 | 1.83 |
| **[1]** | System: 3 x 4 unit cells of graphene,<br>Functional: PBE<br>Basis set: double-zeta w/ polarization orbitals | *unavailable* | 1.86 |
| **[5]** | System: 3 x 3 unit cells of graphene<br>Functional: LDA<br>Basis set: planewave | 1.8 | 1.86* |
| **[6]** | System: 4 x 4 unit cells of graphene<br>Functional: PBE<br>Basis set: planewave | 0.88 | 1.84 |
| **[6]** | System: 4 x 4 unit cells of graphene<br>Functional: PBE + D2<br>Basis set: planewave | 1.04 | 1.84 |
| **[6]** | System: 4 x 4 unit cells of graphene<br>Functional: PBE + D3<br>Basis set: planewave | 0.96 | 1.85 |
| **[6]** | System: 4 x 4 unit cells of graphene<br>Functional: vdW-DF2<br>Basis set: planewave | 0.50 | 1.70 |
| **[3]** | System: 9 x 6 unit cells of graphene<br>Functional: PBE<br>Basis set: planewave | *unavailable* | 1.9* |
| **[2]** | System: 3 x 3 unit cells of graphene<br>Functional: PBE<br>Basis set: planewave | 0.24 | 1.90* |
| **[7]** | System: 5 unit cells of graphene nanoribbon<br>Functional: PBE + vdW-DF<br>Basis set: planewave | *unavailable* | 1.83 |

**Table 2: Adsorption energy and preferred binding site for investigated MD potentials**

| Potential | Site | $E_{ads}$ (eV) |
|---|---|---|
| ReaxFF-1 [11,16] | H | -8.73 |
| ReaxFF-2 [12] | B | -2.18 |
| ReaxFF-3 [13] | H | -5.13 |
| Tersoff-type [10] | H | -0.13 |

**Table 3: Adsorption energies ($E_{ads}$), barriers to adsorption ($E_{bar}$), and barriers to migration ($E_{mig}$) calculated with and without zero-point energy corrections for each of the DFT functionals investigated**

| | Functional | $E_{ads}$ | $E_{ads}$ + ZPE | $E_{bar}$ | $E_{bar}$ + ZPE | $E_{mig}$ | $E_{mig}$ + ZPE |
|---|---|---|---|---|---|---|---|
| C28H14+B | B3LYP | -0.68 | -0.72 | 0.01 | -0.01 | 0.14 | 0.11 |
| | wB97XD | -0.99 | -1.02 | 0.21 | 0.16 | 0.25 | 0.20 |
| | PBE | -1.06 | -1.11 | - | - | - | - |
| C24H12+B | PBE | -0.40 | -0.46 | - | - | 0.09 | 0.09 |

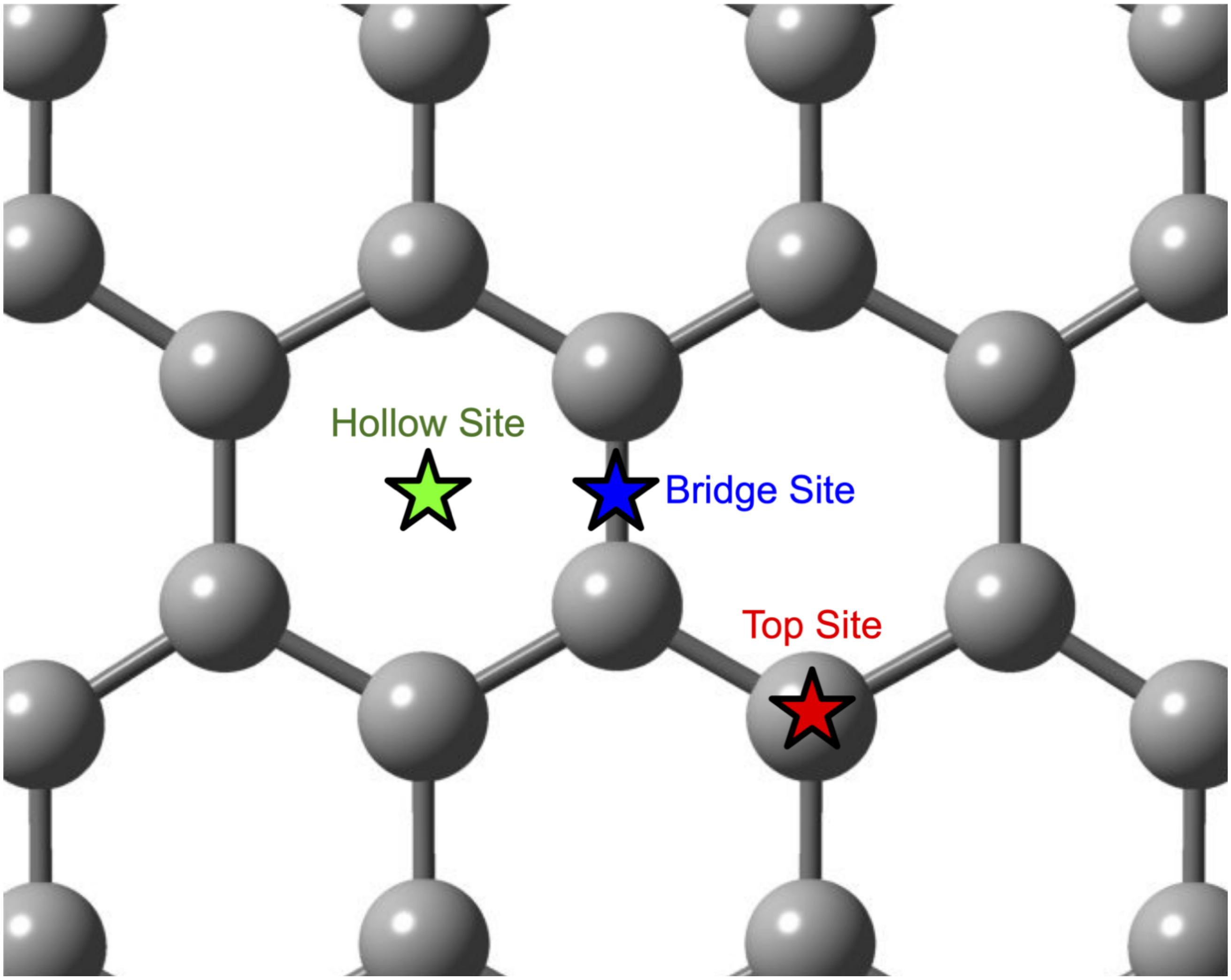

**Figure 1: Diagram of the three possible adatom binding sites on a graphene sheet**

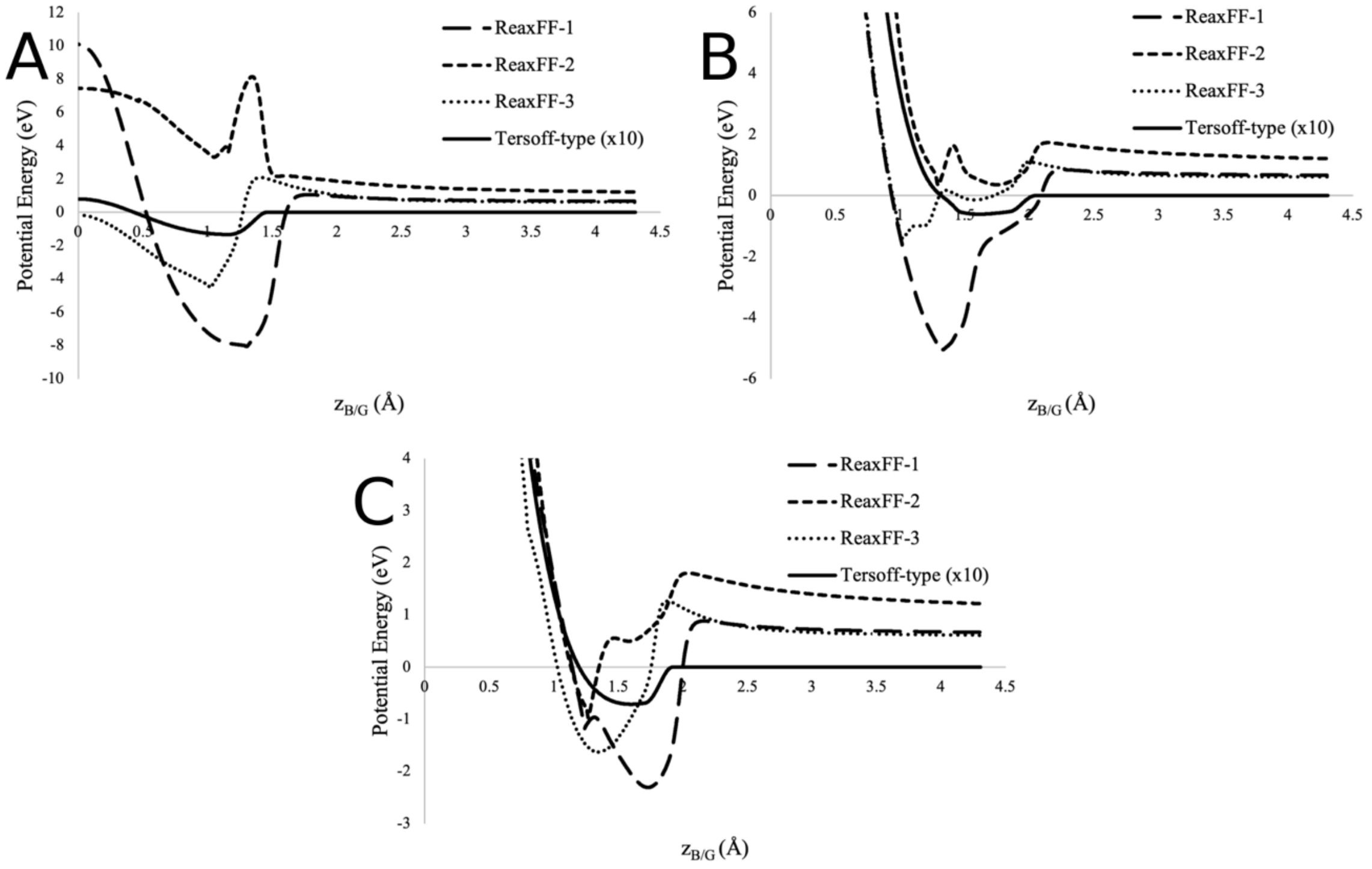

**Figure 2: Potential energy as a function of boron graphene distance ($z_{B/G}$) for the (A) hollow, (B) top, and (C) bridge sites. The energy of the Tersoff-type potential [10] is scaled by a factor of 10 to improve visibility of the minimum in comparison to those of the ReaxFF potentials, ReaxFF-1 [11,16], ReaxFF-2 [12], and ReaxFF-3 [13].**

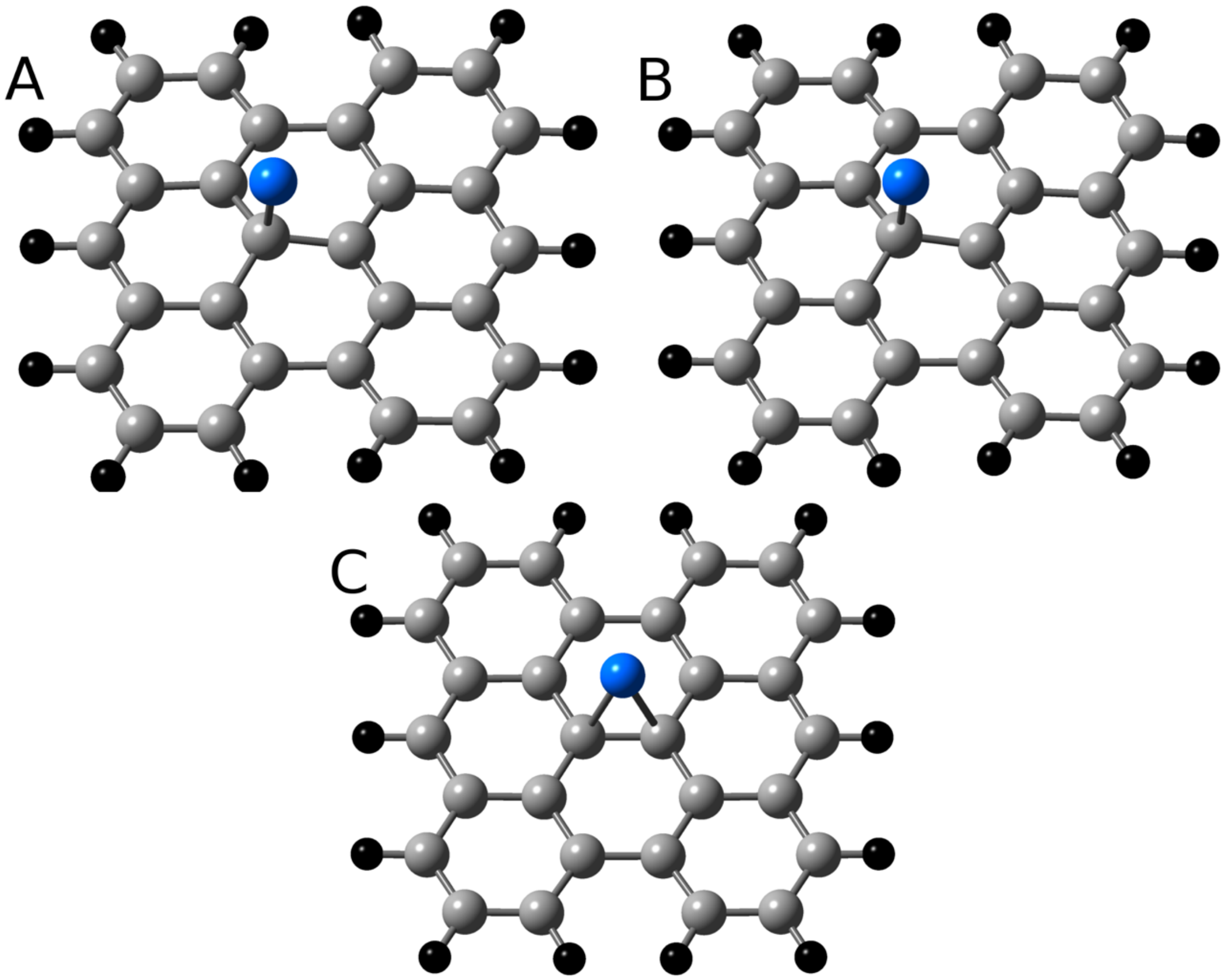

**Figure 3: Binding sites on C28H14 calculated using (A) B3LYP/cc-pVDZ, (B) wB97XD/cc-pVDZ, and (C) PBE/cc-pVDZ methods. Carbon = Grey, Hydrogen = Black, and Boron = Blue.**

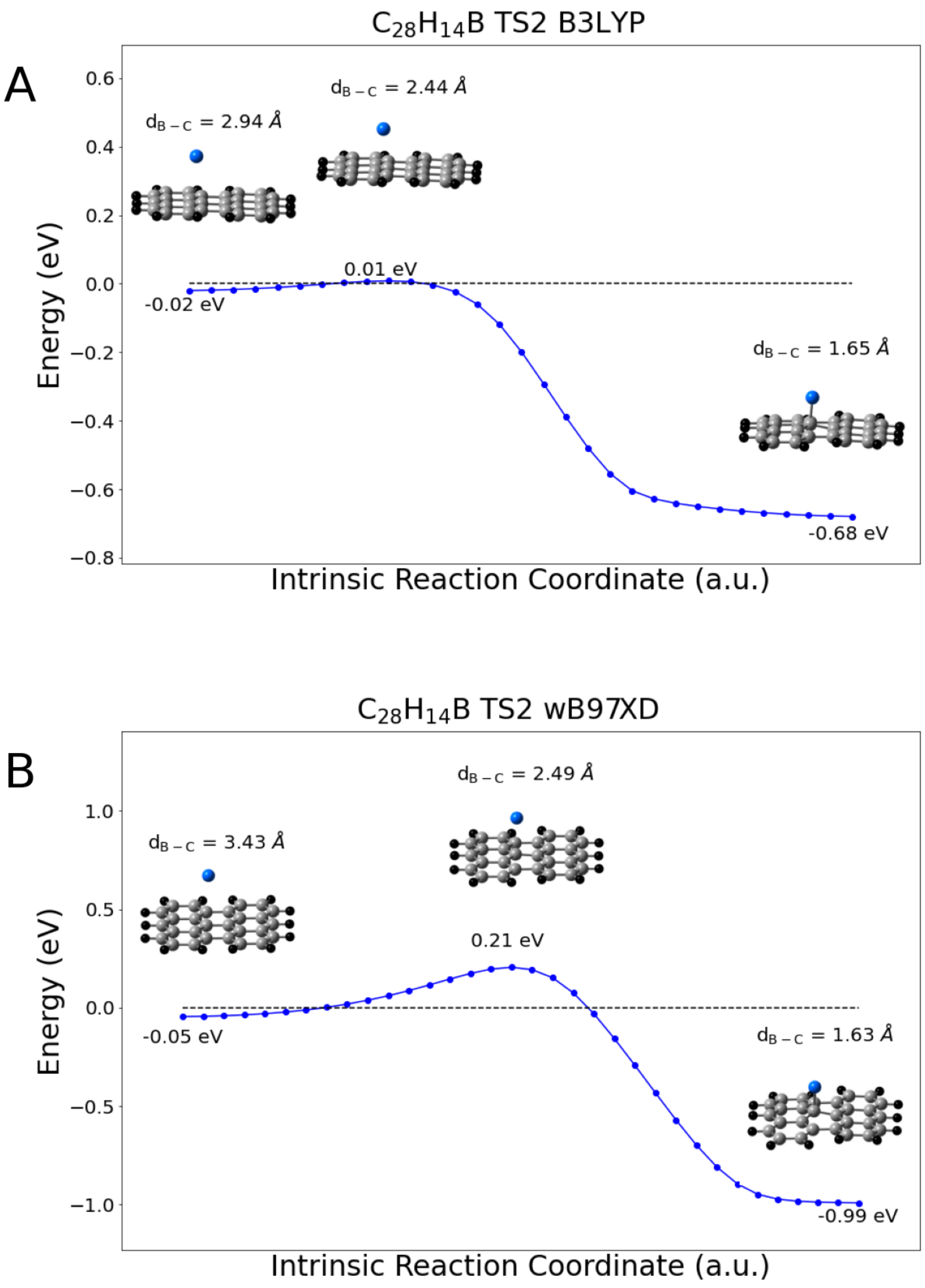

**Figure 4: IRC scans following the reaction from the weakly physisorbed state (left) over the transition state (center) to the fully adsorbed state (right) calculated with the (A) B3LYP/cc-pVDZ method and (B) wB97XD/cc-pVDZ method. The energies depicted are in reference to the energy of the fully separated products, and no zero-point energy corrections have been made. Carbon = Grey, Hydrogen = Black, and Boron = Blue.**

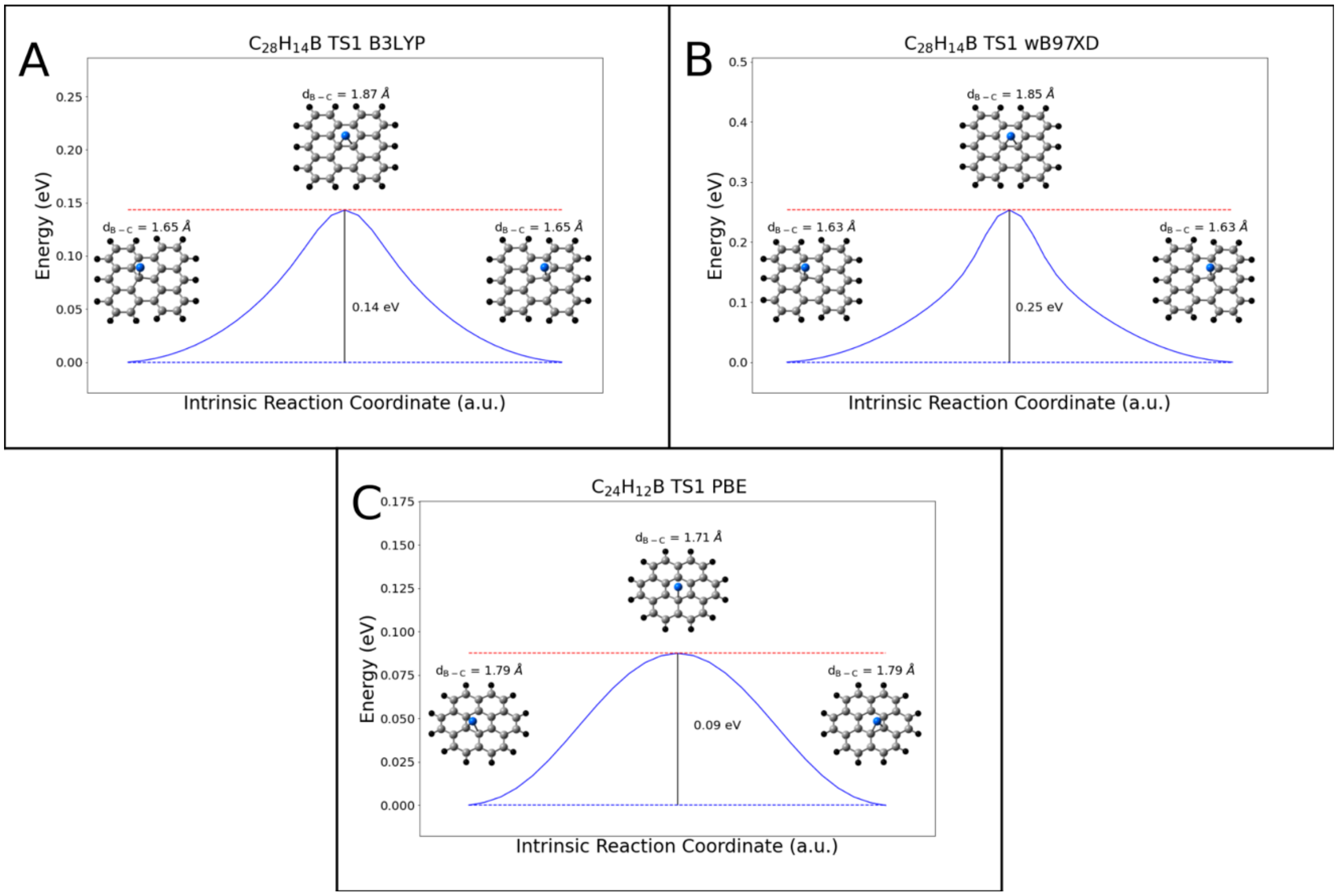

**Figure 5: IRC potential energy scans for the migration of the boron adatom between adjacent binding sites, using the (A) B3LYP functional, (B) wB97XD functional, and (C) PBE functional. Scans performed in (A) and (B) used a bisanthene molecule as a model for graphene, the scan performed in (C) used coronene. The energies depicted are in reference to the energy of the fully bound state, and no zero-point energy corrections have been made. Carbon = Grey, Hydrogen = Black, and Boron = Blue.**

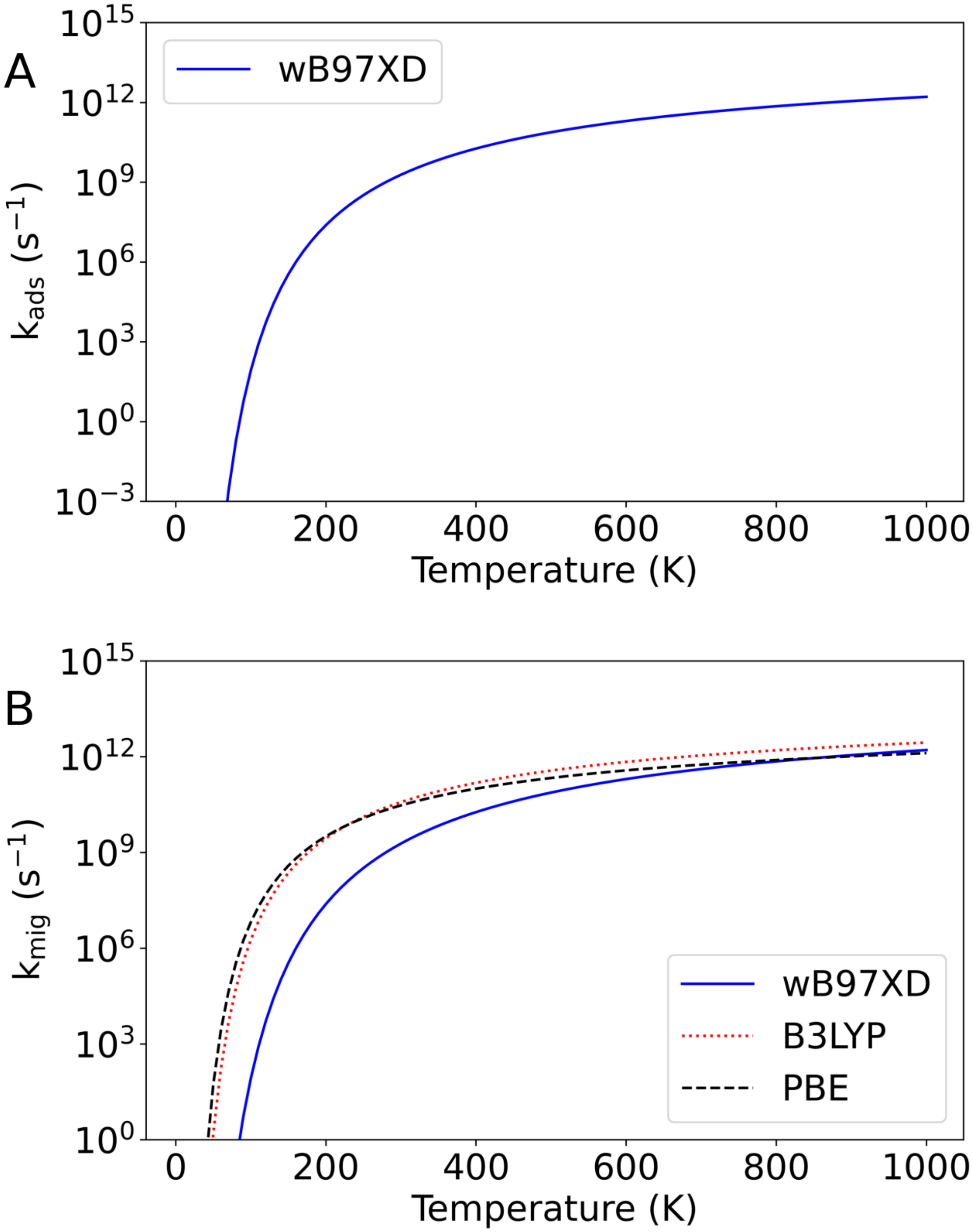

**Figure 6: Reaction rates as a function of temperature for the processes of (A) boron adatom adsorption on graphene and (B) migration of the boron adatom between adjacent binding sites on the graphene surface**

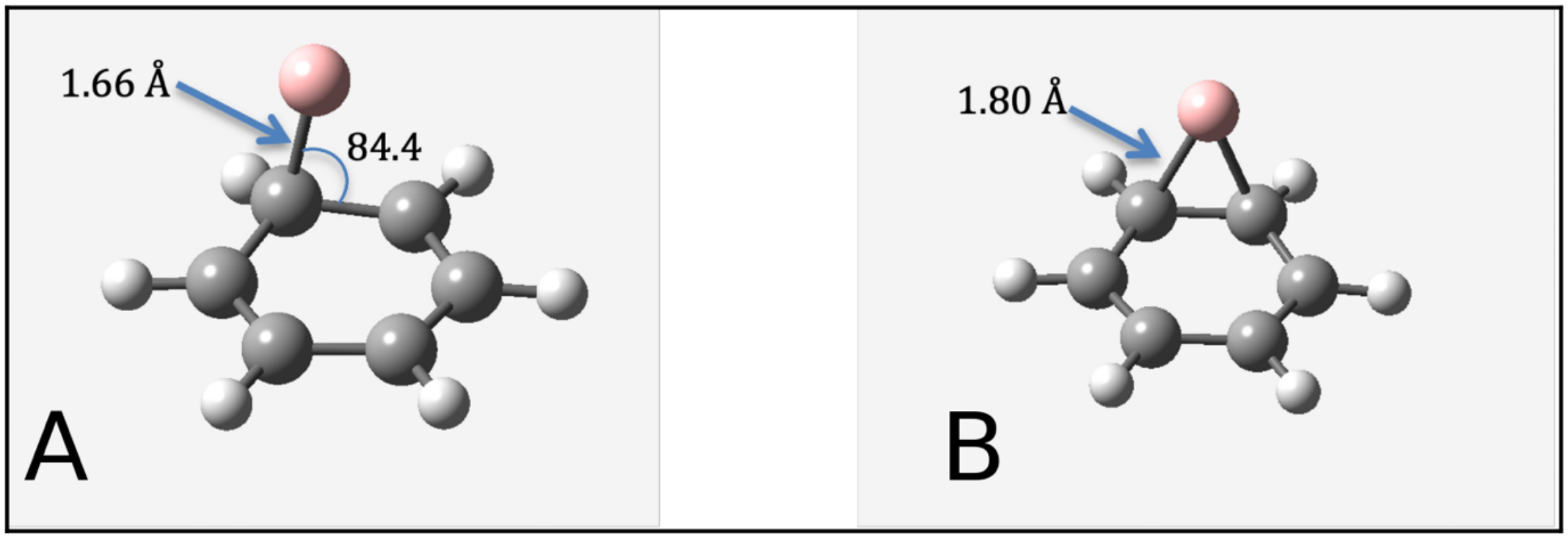

**Figure 7: Stable geometries of atomic boron interacting with benzene as predicted by (A) B3LYP/cc-pVDZ calculations and (B) PBE/cc-pVDZ calculations. Carbon = Grey, Hydrogen = White, and Boron = Pink.**